\documentclass{rspublic}
\usepackage{graphicx}

\newcommand{\vv}[1]{\mathbf{#1}}
\newcommand{\etal}{\textit{et al.\ }}

\begin{document}

\title[Delayed-Feedback Nonlinear Oscillators]{Complex Dynamics and Synchronization of Delayed-Feedback Nonlinear Oscillators}

\author[T. E. Murphy and others]{Thomas E. Murphy$^{1,2}$,
Adam B. Cohen$^{2,3}$,
Bhargava Ravoori$^{2,3}$,
Karl R. B. Schmitt$^{2,4}$,
Anurag V. Setty$^{5}$,
Francesco Sorrentino$^{6}$,
Caitlin R. S. Williams$^{2,3}$,
Edward Ott$^{1,2,3}$ and
Rajarshi Roy$^{2,3,5}$}
\affiliation{
$^1$ Dept. of Electrical \& Computer Engineering,\\
$^2$ Institute for Research in Electronics \& Applied Physics,\\
$^3$ Dept. of Physics,
$^4$ Dept. of Mathematics,\\
$^5$ Institute for Physical Science \& Technology\\University of Maryland, College Park, MD 20742 USA \\
$^6$ Universit\`a degli Studi di Napoli Parthenope, 80143 Napoli, Italy
}

\label{firstpage}

\maketitle

\begin{abstract}{nonlinear dynamics, chaos, synchronization, control}

We describe a flexible and modular delayed-feedback nonlinear oscillator that is capable of generating a wide range of dynamical behaviours, from periodic oscillations to high-dimensional chaos.  The oscillator uses electrooptic modulation and fibre-optic transmission, with feedback and filtering implemented through real-time digital-signal processing.  We consider two such oscillators that are coupled to one another, and we identify the conditions under which they will synchronize.  By examining the rates of divergence or convergence between two coupled oscillators, we quantify the maximum Lyapunov exponents or transverse Lyapunov exponents of the system, and we present an experimental method to determine these rates that does not require a mathematical model of the system.  Finally, we demonstrate a new adaptive control method that keeps two oscillators synchronized even when the coupling between them is changing unpredictably.

\end{abstract}

\section{Introduction}
\label{sec:intro}

Private communications, fast physical random number generators and spatiotemporally distributed sensor networks have provided the context for possible new applications of chaotic dynamical systems.  A key requirement for such applications is the development of reliable and robust generators of chaotic waveforms with broad spectral bandwidths.  By \emph{reliable}, we mean that given parameters of the system, the dynamical properties are reproducible, both experimentally and theoretically.  \emph{Robust} implies that the system exhibits chaotic behaviour over a region of parameter space (i.e., with few periodic windows.)  There have been several realizations of delayed-feedback optoelectronic oscillators that meet these criteria.  Systems that can be configured for integrated optoelectronic fabrication and can function at frequency ranges from tens of GHz down to kHz may find applications in acoustic, biological, chemical, electromagnetic, and mechanical scenarios on nano- to macroscopic spatial scales (Argyris \etal 2005, Kouomou \etal 2005, Uchida \etal 2005, Uchida \etal 2008, Illing \etal 2007, Reidler \etal 2009, Sorrentino \& Ott 2008, 2009, Cohen \etal 2008, Yousefi et al. 2008, Argyris \etal 2008).

Our focus in this paper is on the dynamics of delayed-feedback nonlinear oscillators constructed from modular optoelectronic components.  Delayed feedback enables such systems to generate a wide variety of waveforms, with differing degrees of complexity that depend on the parameters used.  In particular, the time-delay, feedback strength, and filter parameters can be tuned to produce highly stable periodic waveforms (Yao \& Maleki 1996) as well as complex waveforms that are characteristic of robust, high-dimensional chaos (Peil \etal 2009).

In \S\ref{sec:oscillator} we introduce the basic optoelectronic system and the delay-differential equations used for a continuous-time description of the dynamics (Kouomou \etal 2005, Cohen \etal 2008).  The intrinsic nonlinearity of the system arises from the integrated optical Mach-Zehnder modulator which changes the intensity of light transmitted depending on the cosine squared of a modulation voltage applied to its electrodes.   We then chart the dynamical behaviour of the system, using bifurcation diagrams, as the feedback strength and delay time of the feedback loop are varied.   The complexity of the waveforms generated is assessed by the Lyapunov dimension, and we illustrate the wide range of dynamics accessible.

In \S\ref{sec:dsp} we first motivate and then show how to incorporate digital signal processing (DSP) capabilities in the delayed-feedback system.  This enables precise, real time control of system parameters, such as the time-delay and filter characteristics, in a flexible manner well-suited for applications in communications and sensor networks.  The transition from continuous-time to discrete-time equations is outlined; the system is now governed by finite difference equations that describe its time evolution in terms of the system state as sampled at discrete times by an analogue to digital converter (Toomey \etal 2009).   Even though our DSP implementation was aimed at kHz frequencies, such systems can be extended easily into the GHz range.

The question of isochronal synchronization of these nonlinear oscillators (Fischer \etal 2006, Klein \etal 2006, Rogers-Dakin \etal 2006, Schwartz \& Shaw 2007, Zhou \& Roy 2007, Franz \etal 2008) is central to possible applications in sensor networks (Sorrentino \& Ott 2008, 2009).  We thus consider coupled oscillators next in \S\ref{sec:coupling}, where the many different configurations in which even two oscillators may be coupled are outlined.  We then restrict ourselves to the schemes that we have explored in some detail.   A diffusive-coupling scheme that allows the coupled systems to synchronize and retain the dynamical behaviour of the uncoupled systems is of particular interest.   Several results on the dependence of synchronization error on coupling strength that have been obtained mathematically are verified through numerical simulations and tested experimentally.  In particular, we emphasize that, in the experiments, noise and differences in nominally matched system parameters are unavoidable.  We idenfity parameter regimes for the coupling strength where stable synchronization is observed.

While the steady-state synchronization error is an important quantity to measure with regards to sensor and communications applications, the transients towards synchrony and away from synchrony are important as well, and we study these in \S\ref{sec:sync}.  The time scales for these transients set the limits on communication rates and detection of environmental perturbations.  One may determine the maximum Lyapunov exponent for a dynamical system by measurement of transients away from synchrony (Cohen \etal 2008).   When a mathematical model is available, one may predict the dynamics of an experimental system for several delay times by performing data assimilation of experimental data using synchronization of the mathematical model to the data (Cohen \etal 2008, Marino \etal 2009, Quinn \etal 2009, So \etal 1994, Sorrentino \& Ott, 2009b).  Further, it is possible to estimate distributions for  finite-time Lyapunov exponents of  the system.  It should be noted that given two replicas of a dynamical system, one may estimate Lyapunov exponents from transients even when one does not have a mathematical model of the system.

For applications of synchronized chaotic systems to sensor networks a novel adaptive synchronization approach has been recently conceived by Sorrentino \& Ott (2008, 2009).  When the coupling channels between diffusively coupled chaotic dynamical systems serving as nodes of the network are perturbed at time scales slow compared to those of the chaotic fluctuations, they showed that it is possible to not only maintain synchrony between the systems.  In the process of doing so, it is also possible to estimate and track the time-varying perturbations of the coupling strengths.  In the illustrative case of two coupled systems, we have recently shown (Ravoori \etal 2009) that this scheme can be implemented experimentally.  In \S\ref{sec:adaptive} we describe the scheme as implemented in the DSP based system described in \S\ref{sec:dsp}, and we examine its effectiveness in maintaining synchrony and tracking the time-dependent perturbations of the coupling channel.

In \S\ref{sec:conclusion} we summarize our results and discuss future directions of research.

\section{Chaotic optoelectronic oscillator}
\label{sec:oscillator}

\begin{figure}[htbp]
  \centering
  \includegraphics{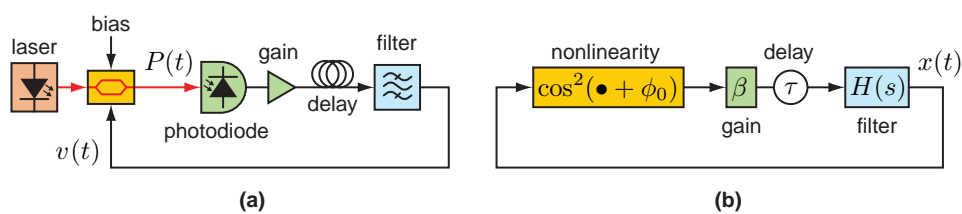}\\
  \caption{Experimental setup and corresponding mathematical block diagram of chaotic optoelectronic oscillator.}\label{fig:system-diagram}
\end{figure}

Figure~\ref{fig:system-diagram}(a) shows a diagram of the chaotic optoelectronic oscillator considered here, composed of a laser, electrooptic intensity modulator, photoreceiver and electrical filter, all connected together in a time-delayed feedback loop.  This system was originally considered by Neyer and Voges (1986), who recognized its potential for bistability and chaos.  The system was later adapted for use as a high-quality microwave oscillator, by incorporating a narrow electrical bandpass filter (Yao \& Maleki 1996).  More recently, there has been renewed interest in using this architecture as a means for generating high-dimensional chaotic waveforms (Kouomou \etal 2005).

The electrooptic modulator is a commercially-available lithium-niobate Mach-Zehnder modulator, identical to those commonly used in optical telecommunication systems.  The input is a continuous-wave optical signal from a distributed feedback laser, which is split into two separate waveguide paths and then recombined, forming an interferometer.  A voltage applied to the modulator induces an optical phase shift between two arms of the interferometer through the linear electrooptic effect.  When the optical signals recombine, the degree to which they interfere constructively depends on the applied voltage.  The optical power emerging from the modulator is then described by (Heismann \etal 1997):
\begin{equation}\label{eq:c1}
    P(t) = P_0 \cos^2\left(\frac{\pi}{2}\frac{v(t)}{V_\pi} + \phi_0\right)\,,
\end{equation}
where $P_0$ is the continuous-wave optical power entering the modulator, $v(t)$ is the voltage applied to the modulator electrodes, $V_\pi$ is the `half-wave voltage', or the voltage required to produce a relative phase shift of $\pi$ between the arms of the interferometer, and $\phi_0$ is an angle describing the bias point of the modulator.  The bias point is controlled either by intentionally making one arm of the interferometer longer or by adding a DC offset to the applied voltage $v(t)$.  The modulators described in this work had a half-wave voltage of $V_\pi = 5.7$ V and were operated at a bias point of $\phi_0 = -\pi/4$.

The modulator converts the applied voltage $v(t)$ into an optical intensity modulation $P(t)$, through the nonlinear modulation function given in equation (\ref{eq:c1}).  We note that this $\cos^2(\bullet)$ modulation function applies to several other optical modulator structures, including liquid crystal modulators, Pockels cells (Hopf \etal 1982), and acoustooptic modulators (Val\'ee \& Delisle 1985).  The same nonlinearity can also be achieved by transmitting an electrically tunable laser through an optical filter that has periodic spectral transmission, such as a single-stage birefringent filter (Goedgebuer 1998) or any other single-pass interferometric filter (Blakely \etal 2004).

The photoreceiver and transimpedance amplifier produce an output voltage $v_{\rm out}(t)$ that is proportional to the optical power $P(t)$,
\begin{equation}\label{eq:c2}
    v_{\rm out}(t) = RGP(t)\,,
\end{equation}
where $R$ is the responsivity of the photodiode (with units of A/W) and $G$ is the net transimpedance gain of the system (with units of V/A.)

The accompanying block diagram in figure~\ref{fig:system-diagram}(b) shows an equivalent mathematical diagram of the system, including the modulator, photoreceiver, amplifiers, filter, and time-delayed feedback.  To simplify the analysis, the voltage applied to the modulator is expressed in normalized units as
\begin{equation}\label{eq:c7}
    x(t) \equiv \frac{\pi}{2}\frac{v(t)}{V_\pi}
\end{equation}
and we collect all of the remaining proportionality constants into a single dimensionless factor that describes the round-trip gain of the loop,
\begin{equation}\label{eq:c8}
    \beta \equiv \frac{\pi}{2}\frac{RGP_0}{V_\pi}\,.
\end{equation}

In terms of these dimensionless variables, the feedback loop relates the filter input $r(t)$ to the filter output $x(t)$ by the following nonlinear transformation and time delay:
\begin{equation}\label{eq:c9}
    r(t) = \beta\cos^2\left[x(t-\tau) + \phi_0\right]\,.
\end{equation}

For the measurements reported here, the electrical filter is a two-pole bandpass filter that is characterized by the linear transfer function
\begin{equation}\label{eq:c3}
    H(s) = \frac{s\tau_H}{(1+s\tau_L)(1+s\tau_H)}\,,
\end{equation}
where $\tau_L$ and $\tau_H$ are the time constants describing the lowpass and high-pass filters, respectively. In the time domain, a linear filter can be represented by state-space differential equations of the form:
\begin{align}
    \label{eq:c4}
    \frac{d\vv{u}}{dt} &= \vv{A}\vv{u}(t) + \vv{B}r(t) \\
    \label{eq:c5}
    x(t) &= \vv{C}\vv{u}(t) + Dr(t)\,,
\end{align}
where $r(t)$ is the input to the filter, $x(t)$ is the output, $\vv{u}(t)$ is a state vector of the filter system, and $\vv{A}$, $\vv{B}$, $\vv{C}$ and $D$ are matrices that describe the bandpass filter.  For the two-pole bandpass filter described by equation~(\ref{eq:c3}), $\vv{u}(t)$ is a two-dimensional vector and the state space matrices can be expressed as
\begin{equation}\label{eq:c6}
    \vv{A} = \begin{bmatrix}
                     -\left(\dfrac{1}{\tau_L}+\dfrac{1}{\tau_H}\right) & -\dfrac{1}{\tau_L} \\
                     \dfrac{1}{\tau_H} & 0 \\
            \end{bmatrix},\quad
    \vv{B} = \begin{bmatrix}
                     \dfrac{1}{\tau_L} \\
                     \vphantom{\dfrac{1}{\tau_L}}0 \\
                   \end{bmatrix},\quad
    \vv{C} = \begin{bmatrix}
             1 & 0 \\
           \end{bmatrix},\quad
    D = 0\,.
\end{equation}

Combining equations~(\ref{eq:c4}), (\ref{eq:c5}), (\ref{eq:c6}) and (\ref{eq:c9}), the system can be described by the following state-space delay differential equation
\begin{equation}\label{eq:c10}
    \frac{d\vv{u}}{dt} = \vv{A}\vv{u}(t) + \vv{B}\beta\cos^2\left[\vv{C}\vv{u}(t-\tau) + \phi_0\right]\,.
\end{equation}

We note that if the bandpass filter is replaced by a simple lowpass filter, then equation (\ref{eq:c10}) simplifies to a scalar delay differential equation that is equivalent to the classic Ikeda system, originally introduced to describe bistability in optical cavities (Ikeda \& Matsumoto 1987).

\begin{table}
\caption{Experimental Parameters of System}
   \label{tab:1}
   \longcaption{The parameters used in experiments and measurements of the nonlinear chaotic oscillator.  Here we give representative values for $P_0$, $G$, $R$, and $V_\pi$, but in practice the factor $\beta$ (c.f. equation~(\ref{eq:c8})) was measured directly by breaking the loop and measuring the small signal, round-trip AC gain.}
   \begin{center}
   \begin{tabular}{|c|c|c|}
   \hline
   \textbf{Parameter} & \textbf{Value} & \textbf{Unit} \\
   \hline\hline
   $P_0$ & 0--50 & $\mu$W \\
   $R$ & $1.0$ & A/W \\
   $G$ & 800 & V/mA \\
   $V_\pi$ & 5.7 & V \\
   $\beta$ & 0--10 & ---\\
   $\tau_H$ & $1.59$ & ms \\
   $\tau_L$ & $15.9$ & $\mu$s \\
   $(2\pi\tau_H)^{-1}$ & $100$ & Hz \\
   $(2\pi\tau_L)^{-1}$ & $10$ & kHz \\
   $\tau$ & $230$ & $\mu$s \\
   $\phi_0$ & $\pi/4$ & rad \\
   \hline
   \end{tabular}
   \end{center}
\end{table}

Table~\ref{tab:1} lists all of the parameter values used in the experiments and simulations.  To simplify the experimental implementation, we consider here a low-frequency system that operates at audio frequencies, but this system can also be scaled to RF or microwave frequencies (Kouomou \etal 2005, Goedgebuer \etal 2002, Cohen \etal 2008).  In practice the round-trip gain ($\beta$) and time delay ($\tau$) were measured experimentally by interrupting the feedback loop at the input to the modulator and measuring the round-trip small-signal gain and group delay using a vector network analyser.  The gain was controlled by varying the optical power $P_0$ entering the modulator.

\begin{figure}[htbp]
  \centering
  \includegraphics{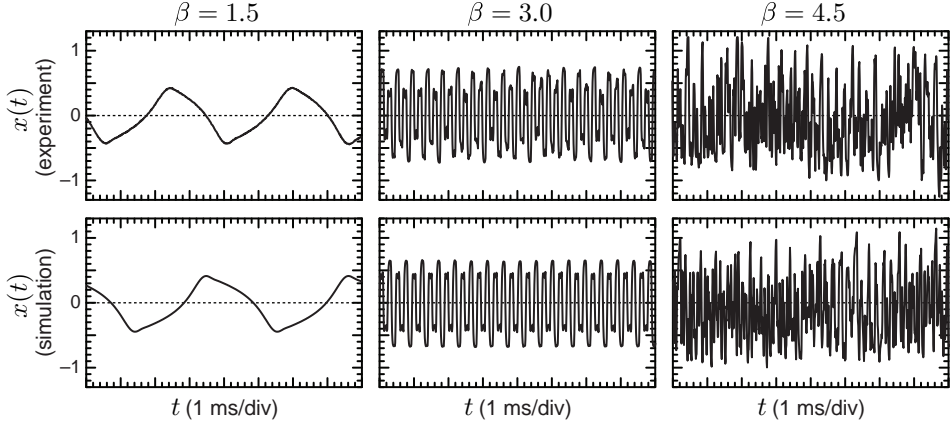}\\
  \caption{Typical measured and calculated time traces for the nonlinear optoelectronic oscillator system, for feedback strengths of $\beta = 1.5$, $3.0$, and $4.5$.}\label{fig:typical-traces}
\end{figure}

\begin{figure}[htbp]
  \centering
  \includegraphics{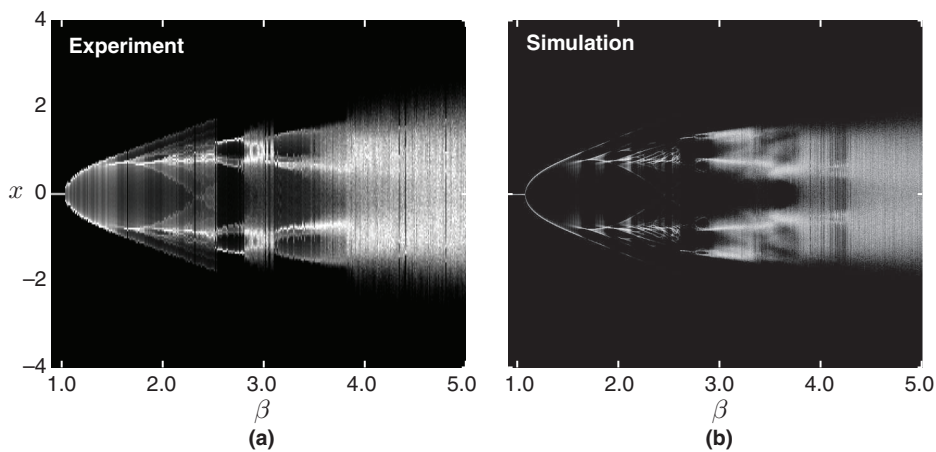}\\
  \caption{Measured and numerically simulated bifurcation diagram, with $\beta$ as an adjustable parameter, for the system parameters given in Table \ref{tab:1}.}\label{fig:bifurcation}
\end{figure}

\begin{figure}[htbp]
  \centering
  \includegraphics{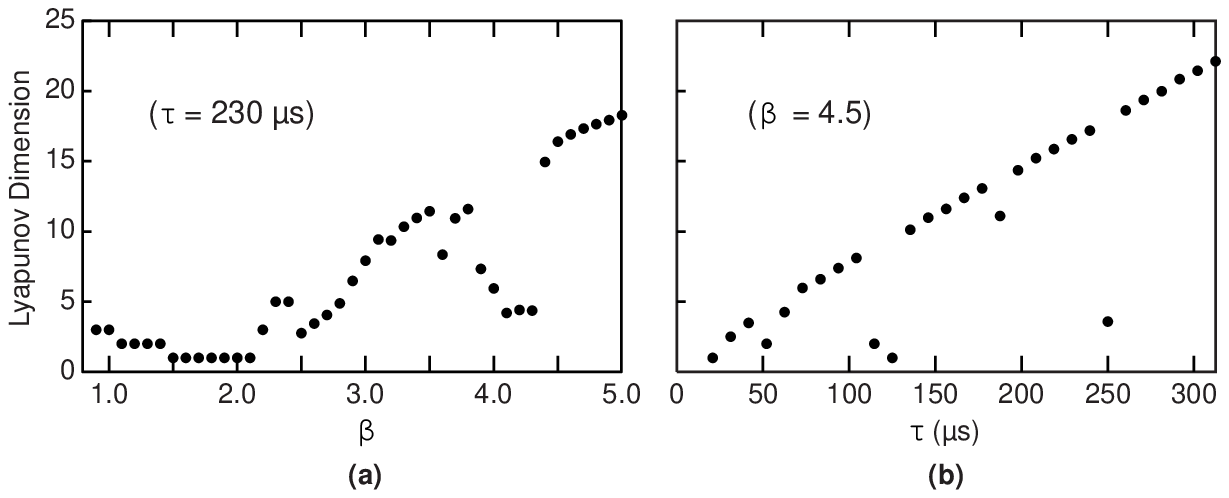}\\
  \caption{Calculated Kaplan-Yorke (Lyapunov) dimension (a) as a function of the feedback strength $\beta$ for fixed feedback delay $\tau$ = 230 $\mu$s and (b) as a function of the the feedback delay $\tau$, for fixed $\beta = 4.5$.  The remaining system parameters are given in Table \ref{tab:1}.}\label{fig:dimensionality}
\end{figure}

In figure~\ref{fig:typical-traces} we show calculated and measured time traces of this system, for three different values of the feedback strength $\beta$, with the time delay and filter parameters given in Table~\ref{tab:1}.  The system exhibits periodic behaviour for small values of $\beta$, but the dynamics become more complex as the feedback strength is increased.  Figure~\ref{fig:bifurcation} plots the measured and simulated bifurcation diagrams with $\beta$ as an adjustable parameter, showing the evolution from periodic to chaotic dynamics.  Peil \etal (2009) reported a detailed experimental and theoretical study of the various regimes of operation of this system.  In Figure~\ref{fig:dimensionality}, we plot the calculated Kaplan-Yorke dimension vs. $\beta$ and vs. $\tau$, showing the progression from simple to high-dimensional chaotic dynamics.  The Kaplan-Yorke dimension (Kaplan \& Yorke 1979) was calculated from the spectrum of Lyapunov exponents, which were numerically computed by solving a linearized version of equation~(\ref{eq:c10}) (Farmer 1982).

\section{Discrete time implementation}
\label{sec:dsp}

The optoelectronic oscillator described in \S\ref{sec:oscillator} was introduced using a continuous-time delay-differential equation, but in practice, we implemented the system using discrete-time digital signal processing (DSP) technology.  DSP provides a flexible platform for programmable filtering and delay operations, and offers a number of advantages over conventional analogue filters, especially when high-speed performance is not required.  For example, it is easy to program two digital filters to have identical characteristics, whereas matching of analogue filters relies on finding identical components such as resistors, capacitors and amplifiers.  DSP systems are especially advantageous in synchronization experiments, where mismatched parameters between nominally identical systems can otherwise impair the synchrony between the two systems.

Analogue delay lines typically use either optical fibre or coaxial cables to achieve a time delay of $L/v$, where $L$ is the length of the transmission medium and $v$ is the propagation speed.  Such systems cannot be scaled to large time delays because the required delay lines are either impractically long or prohibitively lossy.  DSP systems, by contrast, can produce a lossless time delay that is limited only by the available memory and sampling rate.  The use of digital processing in nonlinear dynamical systems dates to 1982, when Hopf \etal used a computer and ADC/DAC to achieve long delay times for a similar optoelectronic oscillator.  Since then, digital signal processing systems have dramatically improved in performance and cost, and are now commonplace in consumer electronics.

Perhaps the most compelling argument in favor of DSP is that it allows real-time adjustment of the the gain, delay and filter coefficients -- parameters that are typically static in analogue filter systems.  In \S\ref{sec:adaptive}, we describe an adaptive control scheme that takes advantage of this flexibility provided by digital processing.

\begin{figure}[htbp]
  \centering
  \includegraphics{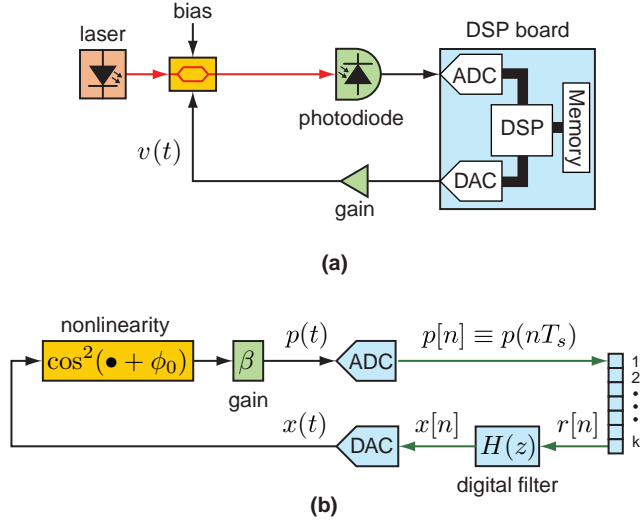}\\
  \caption{(a) Experimental setup showing the use of digital signal processing hardware to implement bandpass filter and delay. (b) Equivalent discrete-time mathematical block diagram of dynamical system.}\label{fig:dsp-setup}
\end{figure}

Figure~\ref{fig:dsp-setup}(a) shows how the original experimental apparatus (figure~\ref{fig:system-diagram}(a)) was adapted to incorporate digital signal processing.  The system uses the same laser, electrooptic modulator and photoreceiver as its continuous-time counterpart, but the filtering and delay are performed using a digital signal processing board.  The DSP board uses an analogue-to-digital converter (ADC) to sample and digitize the input signal $r(t)$, forming a discrete-time input sequence,
\begin{equation}\label{eq:dsp1}
    p[n] \equiv p(nT_s)\,,
\end{equation}
where $n$ is an integer and $T_s$ denotes the sampling period.  The discrete-time signal $p[n]$ is stored in a memory buffer to produce the desired delay and then digitally filtered.  The output signal $x[n]$ is then routed through a complementary digital-to-analogue converter (DAC) to yield the analogue output signal $x(t)$ that drives the electrooptic modulator.  The complete system is therefore a hybrid discrete / continuous-time system that retains the advantages of optical signal transmission, while exploiting the flexibility of discrete-time signal processing.

The DSP board used in these experiments contains a 225 MHz floating-point DSP processor, 64 MB RAM, and a 16-bit ADC/DAC. The maximum sampling frequency was limited by the ADC/DAC chip, which was designed for audio signals.  Except where noted, we used a sampling rate of $1/T_s =$ 96 kS/s in these experiments, although the system could be scaled to higher frequencies by replacing the ADC/DAC hardware.  The lowpass filter in the feedback loop restricts dynamical behaviour to frequencies well below the Nyquist frequency, ensuring that the sampling does not contribute signficantly to the filtering.  Higher-performance field-programmable gate array (FPGA) boards could perform the same operations at sampling rates as high as 1 GS/s.

Figure~\ref{fig:dsp-setup}(b) shows a mathematical block diagram of the discrete-time system.  In this system, $k$ denotes the feedback delay, which we take to be an integer number of timesteps, and the dynamical filter is described by discrete difference equations rather than differential equations.  The digital filter was designed to act as a two-pole bandpass filter that approximates the response of the continuous-time filter described in equations~(\ref{eq:c3})--(\ref{eq:c5}).  The discrete-time transfer function $H(z)$ is obtained from the continuous-time transfer function $H(s)$ by applying a bilinear transform with frequency pre-warping (Oppenheim \etal 1999).  This process yields the following equivalent discrete-time transfer function
\begin{equation}\label{eq:dsp2}
    H(z) = \frac{1}{4}(1-z_L)(1+z_H)\frac{(1-z^{-2})}{(1-z_Lz^{-1})(1-z_Hz^{-1})}\,,
\end{equation}
where $z_L$ and $z_H$ are the poles of the discrete-time filter, which are related to the time constants $\tau_L$ and $\tau_H$ and sampling period $T_s$ by
\begin{equation}\label{eq:dsp3}
    z_H = \frac{1-\tan\left(\dfrac{T_s}{2\tau_H}\right)}{1+\tan\left(\dfrac{T}{2\tau_H}\right)},\qquad
    z_L = \frac{1-\tan\left(\dfrac{T_s}{2\tau_L}\right)}{1+\tan\left(\dfrac{T}{2\tau_L}\right)}\,.
\end{equation}

The discrete-time filter can be represented by state-space evolution equations analogous to equations~(\ref{eq:c4}) and (\ref{eq:c5}),
\begin{align}
    \label{eq:dsp4}
    \vv{u}[n+1] &= \vv{A}\vv{u}[n] + \vv{B}r[n] \\
    \label{eq:dsp5}
    x[n] &= \vv{C}\vv{u}[n] + Dr[n]\,,
\end{align}
where $r[n]$ is the filter input, $x[n]$ is the output, and $\vv{u}[n]$ is a two-dimensional state vector.  For the filter described in equation~(\ref{eq:dsp2}), the state space matrices can be expressed as
\begin{alignat}{3}
    \label{eq:dsp6}
    \vv{A}& = \begin{bmatrix}
                     -(z_L + z_H) & -z_L \\
                     z_H & 0 \\
            \end{bmatrix}&&
    \vv{B} = \begin{bmatrix}
                     z_L \\
                     0 \\
              \end{bmatrix}\\
    \label{eq:dsp7}
    \vv{C}& = \begin{bmatrix}
                0 & -\dfrac{(1-z_L)(1+z_H)(1+z_Lz_H)}{4z_Lz_H} \\
              \end{bmatrix}&\qquad&
    D = \frac{1}{4}(1-z_L)(1+z_H)\,.
\end{alignat}
The filter input is related to the filter output through a nonlinearity and delay,
\begin{equation}\label{eq:dsp8}
    r[n] = \beta\cos^2\left(x[n-k] + \phi_0\right)\,,
\end{equation}
where $\beta$ is the round-trip gain defined in equation~(\ref{eq:c8}) and the delay is chosen to be $k=22$, which, at a sampling rate of 96 kS/s, corresponds to a feedback delay of 830 $\mu$s.

\section{Coupled systems and synchronization}
\label{sec:coupling}

An interesting property of chaotic systems is that two systems, when properly coupled together, can synchronize with one another and evolve along the same chaotic orbit (Fujisaka \& Yamada 1983, Pecora \& Carroll 1990, Pikovsky \etal 2001, Boccaletti 2008).  Many proposed applications of chaos, including secure communication systems, sensor networks, and data assimilation and prediction, rely on this phenomenon of synchronization between chaotic oscillators (Kanter \etal 2008, Argyris \etal 2005, Golubitsky \etal 2005, Boccaletti \etal 2006, Arenas \etal 2008).  There have been some analytical studies of the coupling threshold required for synchronization in delayed-feedback systems (Pyragas 1998, B\"unner \& Just 1998). Peil \etal (2007) reported some experimental measurements and theoretical models of synchronization between time-delayed optoelectronic oscillators like those discussed here.  We seek in the this section to more thoroughly investigate how two such systems can be coupled together, and the conditions under which they can synchronize.

\begin{figure}[htbp]
  \centering
  \includegraphics{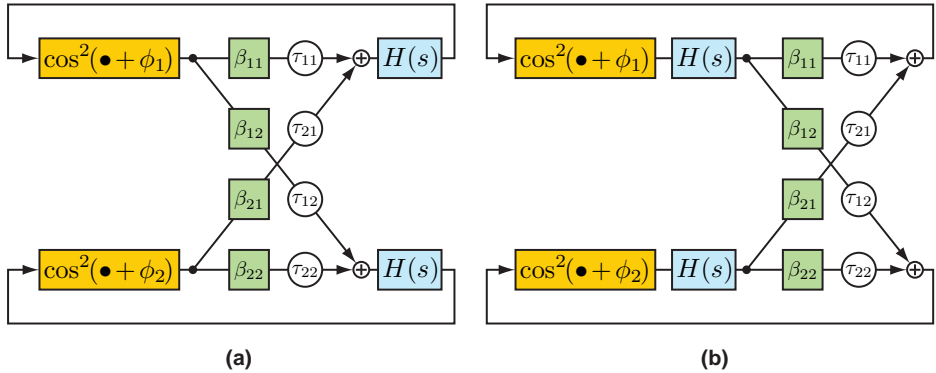}\\
  \caption{(a) Block diagram of two linearly-coupled optoelectronic chaotic oscillators, where the coupling and delays are taken to be in the optical path connecting the two systems.  (b) Equivalent system, obtained by commuting the coupling and delay with the bandpass filter.  In practice, the coupling is implemented optically as in (a), but for convenience, we analyse the equivalent scenario depicted in (b).
  }\label{fig:optical-coupling}
\end{figure}

The block diagram in figure~\ref{fig:optical-coupling}(a) shows the most general type of linear optical coupling between two systems.  In this case, we imagine that the optical signal emerging from the modulator in system 1 is split and fed back into both systems.  The constants $\beta_{11}$ and $\tau_{11}$ denote the self-feedback gain and delay for system 1, and $\beta_{12}$ and $\tau_{12}$ describe the coupling from system $1\rightarrow 2$.  Similarly, $\beta_{22}$, and $\tau_{22}$ are the self-feedback parameters of system 2, and $\beta_{21}$ and $\tau_{21}$ describe the coupling from $2\rightarrow 1$.  We assume that the bandpass filters between the two systems are identical.

The filter ($H(s)$), gain ($\beta_{ij}$) and delay ($\tau_{ij}$) are all linear, time-invariant operations, and they can therefore be freely permuted without changing the dynamics of the system.
Using these arguements, one can transform the optically coupled system shown in figure \ref{fig:optical-coupling}a to the equivalent system shown in figure \ref{fig:optical-coupling}b, where the coupling instead applies to the electrical signals $x_j(t)$ emerging from the bandpass filters.  This coupling configuration can be described by the following coupled delay differential equations:
\begin{align}
    \label{eq:s1}
    \frac{d\vv{u}_1}{dt} &= \vv{A}\vv{u}_1(t) + \vv{B}
    \cos^2\left[\beta_{11}\vv{C}\vv{u}_1(t-\tau_{11}) + \beta_{21}\vv{C}\vv{u}_2(t-\tau_{21}) + \phi_1\right] \\
    \label{eq:s2}
    \frac{d\vv{u}_2}{dt} &= \vv{A}\vv{u}_2(t) + \vv{B}
    \cos^2\left[\beta_{22}\vv{C}\vv{u}_2(t-\tau_{22}) + \beta_{12}\vv{C}\vv{u}_1(t-\tau_{12}) + \phi_2\right]\,,
\end{align}
where $\vv{u}_1$ and $\vv{u}_2$ are the state-vectors for the bandpass filters in oscillators 1 and 2, respectively.

To understand the conditions under which synchrony can occur, we begin by assuming that a synchronous solution exists,
\begin{equation}\label{eq:s3}
    \vv{u}_1(t) = \vv{u}_2(t-\tau_0) \equiv \vv{u}(t)\,,
\end{equation}
where we have allowed for lag synchrony with a time delay $\tau_0$.  Upon substituting this assumption into equations (\ref{eq:s1}) and (\ref{eq:s2}), we obtain self-consistent dynamical equations for $\vv{u}(t)$ only under the following conditions:
\begin{equation}\label{eq:s4}
    \phi_1 = \phi_2,\quad
    \beta_{11} = \beta_{22},\quad
    \beta_{12} = \beta_{21},\quad
    \tau_{11} = \tau_{22},\quad
    \tau_0 = \frac{1}{2}(\tau_{21} - \tau_{12})\,.
\end{equation}
While these conditions are necessary for a synchronous solution to exist, they do not guarantee the stability of this solution.

We now further restrict our attention to cases in which the systems synchronize in a state that obeys the same dynamical equation as that of an uncoupled, isolated system described by parameters $\beta$, $\tau$ and $\phi_0$.  This lifts the constraint that $\beta_{11} = \beta_{22}$ and $\beta_{12} = \beta_{21}$, but imposes the following additional conditions for synchrony:
\begin{align}
    \label{eq:s5}
    \beta_{11} + \beta_{21} = \beta_{22} + \beta_{12} &\equiv \beta \\
    \label{eq:s6}
    \tau_{11} = \tau_{21} = \tau_{22} = \tau_{12} &\equiv \tau \\
    \label{eq:s7}
    \phi_1 = \phi_2 &\equiv \phi_0\\
    \tau_0 = 0\,.
\end{align}
In this scenario, which is termed `diffusive coupling', the constraint on the coupling conditions (equation (\ref{eq:s5})) can be cast in terms of two dimensionless parameters $\kappa_1$ and $\kappa_2$, defined through the relations
\begin{gather}
    \label{eq:s8}
    \beta_{21} = \kappa_1\beta,\qquad \beta_{11} = (1-\kappa_1)\beta\,\, \\
    \label{eq:s9}
    \beta_{12} = \kappa_2\beta,\qquad \beta_{22} = (1-\kappa_2)\beta\,.
\end{gather}
With this definition, $(1-\kappa_1)$ and $\kappa_1$ describe relative proportions of self-feedback vs. cross-coupled feedback, respectively, entering system 1, and $\kappa_2$ has a similar interpretation for system 2.

\begin{figure}[htbp]
  \centering
  \includegraphics{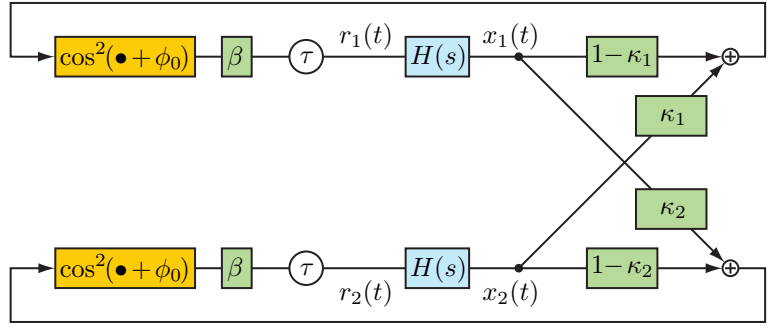}\\
  \caption{Block diagram of two diffusively-coupled oscillators.  The coupling is constructed in a way that guarantees that the resulting system admits a synchronous solution of the form $x_1(t) = x_2(t)\equiv x(t)$ where $x(t)$ exhibits the same dynamical behaviour as that of an isolated system.}\label{fig:diffusive-coupling}
\end{figure}

Figure \ref{fig:diffusive-coupling} presents the block diagram of two diffusively-coupled oscillators.  In order to make the equations comparable in form to the single-oscillator system described in \S\ref{sec:oscillator}, we have factored out a common scale factor $\beta$ from all four of the coupling terms and commuted this scale factor with the bandpass filter $H(s)$.

The diffusively coupled oscillator system shown in figure \ref{fig:diffusive-coupling} is described by the following coupled equations
\begin{align}
    \label{eq:s10}
    \frac{d\vv{u}_1}{dt} &= \vv{A}\vv{u}_1(t) + \vv{B}\beta
    \cos^2\Bigl(\vv{C}\bigl[(1-\kappa_1)\vv{u}_1(t-\tau) + \kappa_1\vv{u}_2(t-\tau)\bigr] + \phi_0\Bigr) \\
    \label{eq:s11}
    \frac{d\vv{u}_2}{dt} &= \vv{A}\vv{u}_2(t) + \vv{B}\beta
    \cos^2\Bigl(\vv{C}\bigl[(1-\kappa_2)\vv{u}_2(t-\tau) + \kappa_2\vv{u}_1(t-\tau)\bigr] + \phi_0\Bigr)\,.
\end{align}

These equations can be seen to admit an isochronally synchronized solution that, when synchronized, satisfies the same equation (\ref{eq:c10}) given earlier for an isolated system.

To investigate the stability of the synchronized solution, we perform following change of variables
\begin{equation}
    \label{eq:s12}
    \vv{u}_+(t) = \frac{1}{2}\left[\vv{u}_1(t) + \vv{u}_2(t)\right],\qquad
    \vv{u}_-(t) = \frac{1}{2}\left[\vv{u}_1(t) - \vv{u}_2(t)\right]\,,
\end{equation}
where the difference $\vv{u}_-(t)$ is expected to converge to zero for a stable synchronous solution. Expressing equations (\ref{eq:s10}) and (\ref{eq:s11}) in terms of the $\vv{u}_\pm$, and linearizing about the synchronous state, we find
\begin{align}
    \label{eq:s13}
    \frac{d\vv{u}_+}{dt} &= \vv{A}\vv{u}_+(t) + \vv{B}\beta
    \cos^2\Bigl(\vv{C}\vv{u}_+(t-\tau) + \phi_0\Bigr) \\
    \label{eq:s14}
    \frac{d\vv{u}_-}{dt} &= \vv{A}\vv{u}_-(t) + \vv{B}\beta
    \sin\Bigl(2\vv{C}\vv{u}_+(t-\tau) + 2\phi_0\Bigr)
    (\kappa_1+\kappa_2 - 1)\vv{u}_-(t-\tau)\,.
\end{align}

Comparing equations (\ref{eq:s13}) and (\ref{eq:c10}), we see that $\vv{u}_+(t)$ satisfies the same dynamical equation as isolated system, as expected.  The two coupling parameters appear in the second equation only in the combination $(\kappa_1 + \kappa_2)$.  We therefore conclude that for a given $\beta$, $\tau$ and $\phi_0$, the stability of the synchronous solution depends only on the sum $(\kappa_1 + \kappa_2)$, but not on the values of $\kappa_1$ and $\kappa_2$ individually.

Furthermore, in the special case that $\kappa_1+\kappa_2 = 1$, equation (\ref{eq:s14}) simplifies to
\begin{equation}
    \label{eq:s15}
    \frac{d\vv{u}_-}{dt} = \vv{A}\vv{u}_-(t)\,.
\end{equation}
Because the linear bandpass filter is stable (i.e., $\vv{A}$ has negative eigenvalues) the difference vector $\vv{u}_-(t)$ will always converge to zero according to the filter time-constants $\tau_L$ and $\tau_H$ whenever $\kappa_1+\kappa_2 = 1$.

\begin{figure}[htbp]
  \centering
  \includegraphics{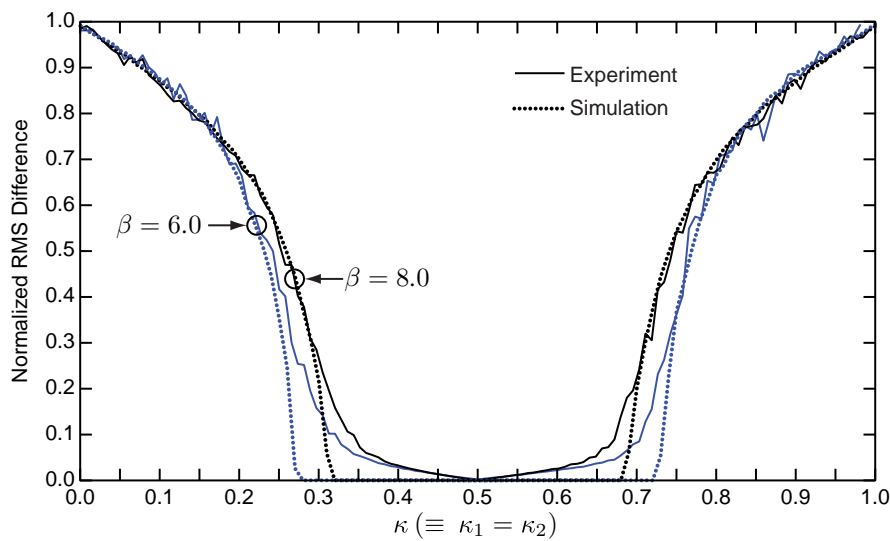}\\
  \caption{Measured and simulated normalized synchronization error as a function of $\kappa$ for $\beta = 6$ and $\beta = 8$.  The measurements and simulations were conducted using symmetric bidirectional coupling, $(\kappa_1 = \kappa_2 \equiv \kappa)$.}\label{fig:sync-error}
\end{figure}

Figure \ref{fig:sync-error} plots the measured and simulated normalized root-mean square (RMS) synchronization error as a function of the coupling strength $\kappa$ for the case of bidirectional symmetric coupling, i.e., $\kappa_1 = \kappa_2 \equiv \kappa$, showing the regimes in which the two systems synchronize.  We define the normalized synchronization error as
\begin{equation}\label{eq:s16}
    \sigma_x \equiv \left(\frac{\left<(x_1(t) - x_2(t))^2\right>}{\left<x_1^2(t) + x_2^2(t)\right>}\right)^{1/2}\,,
\end{equation}
where $\left<\bullet\right>$ indicates a time average.  The normalized error $\sigma_x$ is zero in the case of a synchronized solution, but approaches 1 in the limit that the two signals are identically distributed, but uncorrelated.  While the experimental measurements and simulations were performed by taking $\kappa_1 = \kappa_2$, the results can be generalized to other combinations of $\kappa_1$ and $\kappa_2$ because the synchronization condition depends only on the sum $\kappa_1+\kappa_2$.

As indicated in figure \ref{fig:sync-error}, the two systems synchronize unconditionally for the special case that $\kappa_1 + \kappa_2 = 1$, and they synchronize for a range of $(\kappa_1 + \kappa_2)$ centered symmetrically about this point.  The range of values over which the system synchronize is found to depend on the feedback gain $\beta$.  In general, we observed that the higher values of $\beta$ (and higher Lyapunov dimension) yield a narrower synchronization regime.

\section{Synchronization -- Transient dynamics}
\label{sec:sync}

In addition to knowing \emph{whether} two systems synchronize, it is also important to understand the \emph{rate} at which they converge to a synchronous state, which is quantified by the transverse Lyapunov exponent (Fujisaka \& Yamada 1983, Pecora \& Carroll 1998).  The transverse Lyapunov exponent (TLE), denoted $\lambda_T$, defines the average exponential rate at which a pair of coupled identical oscillators converge or diverge in phase space.  A negative TLE corresponds to converging trajectories, indicating stable synchronization, while a positive exponent indicates diverging trajectories that do not synchronize.

The TLE defines an important timescale in applications such as chaotic communication and sensor networks that rely on synchronization.  In chaotic sensor networks with time-varying coupling, the TLE limits the speed of perturbations that the system can track.  In a chaotic communication system, the TLE limits the attainable bit-rate that can be successfully decoded.

Equally important is the (positive) maximal Lyapunov exponent, which describes the rate at which initially synchronous solutions diverge from one another when they are decoupled.  This divergence rate is important in data assimilation and prediction applications, which use synchronization to predict the future behaviour of a dynamical system.  Here we present a numerical and experimental study of the transient synchronization and desynchronization dynamics of two coupled chaotic optoelectronic oscillators.

One method to determine the transverse Lyapunov exponent is to suddenly couple two independent and identical chaotic oscillators.  By analysing the transition from the initially uncorrelated dynamics to a synchronous state, we can determine the (finite time) transverse Lyapunov exponent of the system.  Conversely, if the two systems are initially synchronized, the coupling can be suddenly turned off, allowing the trajectories to exponentially diverge.  By measuring the rate of exponential divergence, we find the maximal Lyapunov exponent of the system (Cohen \etal 2008).  Unlike conventional methods, which require numerical solution of a linearized system of equations, this approach can be applied even in cases when an exact model of the physical system is unavailable or impractical.  As long as two experimental systems can be made to synchronize, the Lyapunov exponents describing synchronization and desynchronization can be determined from transient time-series analysis.

\begin{figure}[htbp]
  \centering
  \includegraphics{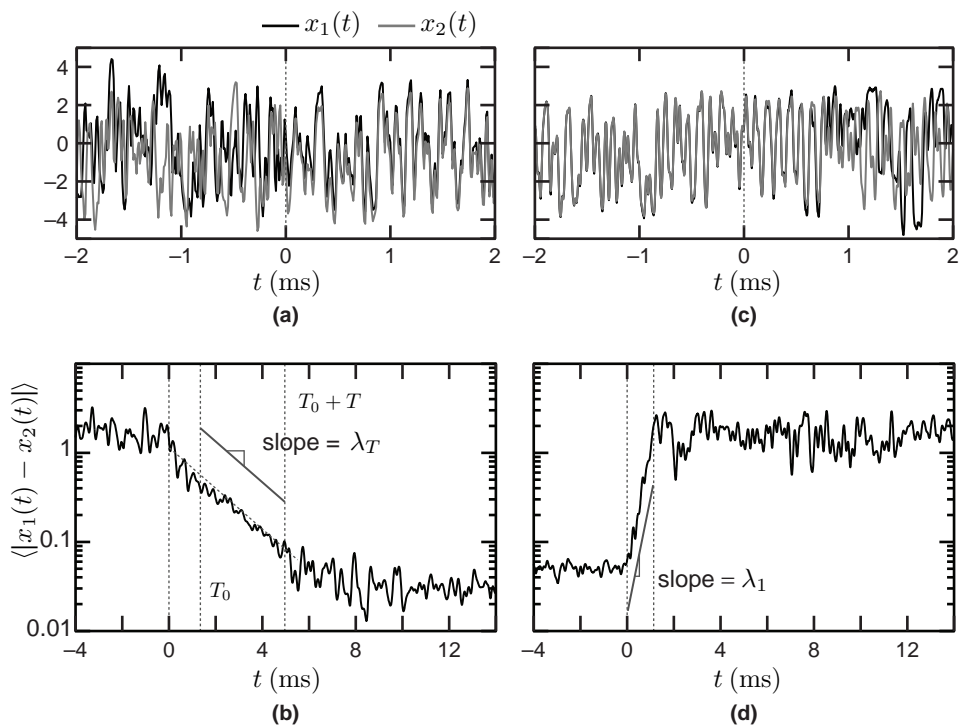}\\
  \caption{(a) Experimentally measured time series showing synchronization of two coupled chaotic oscillators. The two systems were uncoupled for $t<0$ and symmetric bidirection coupling was abruptly enabled at $t=0$. (b) Measured absolute difference $|x_1(t) - x_2(t)|$ plotted on a logarithmic scale, and smoothed to show exponential convergence of trajectories. By fitting a line to this slope, one can estimate the finite-time transverse Lyapunov exponent $\lambda_T$, which characterizes the timescale over which synchronization occurs. (c) Experimentally measured time series showing divergence of two initially synchronized systems, when the coupling is disabled at $t=0$. (d) The finite-time maximal Lyapunov exponent $\lambda_1$ is estimated by measuring the average exponential divergence rate.}\label{fig:convergence}
\end{figure}

This method of determining the Lyapunov exponent is illustrated in Figure~\ref{fig:convergence}, which shows the exponential convergence and divergence of two coupled chaotic optoelectronic oscillators.  In figure \ref{fig:convergence}(a)-(b), the two oscillators were initially uncoupled for $t<0$, but the coupling was suddenly enabled at $t=0$.  Specifically, for $t>0$ the systems were bidirectionally coupled as shown in figure~\ref{fig:diffusive-coupling} with $\kappa_1 = \kappa_2 \equiv \kappa = 0.4375$.  Figure~\ref{fig:convergence}(a) plots the measured outputs $x_1(t)$ and $x_2(t)$ for one representative case, showing the transition from uncorrelated to synchronized dynamics.  Figure~\ref{fig:convergence}(b) shows the absolute difference $|x_1(t)-x_2(t)|$, smoothed with a 100 $\mu$s sliding window average, and plotted on semilogarithmic axes to clearly show the exponential convergence.  By fitting an exponential relation to this curve, we determine the (negative) transverse Lyapunov exponent.  Figure~\ref{fig:convergence}(c)-(d) show similar data obtained when two initially synchronized systems are decoupled at $t=0$, allowing them to exponentially diverge.  In this case, the (positive) maximum Lyapunov exponent $\lambda_1$ is similarly determined by finding the best-fitted slope to the smoothed logarithmic difference between the two traces.

When determining the Lyapunov exponent using this method, the exponential convergence or divergence is estimated only over a finite fitting interval $T$.  In practice, the allowable fitting interval is restricted by the synchronization error floor, which is caused by noise and mismatches between the two systems (Shahverdiev \etal 2005).  In numerical simulations, the convergence/divergence can be observed over many orders of magnitude, and we can therefore fit the exponential relation over a larger time window $T$.  If the fitting window is long enough to span the entire chaotic attractor, this calculation reveals the `global' or `asymptotic' Lyapunov exponent.  For a short fitting interval, the trajectory remains only in a localized portion of the chaotic attractor, and thus we obtain only a `local' Lyaponov exponent.  The local Lyapunov exponents vary about an attractor, and their statistical distribution depends upon the dynamical nature of the coupled system.

\begin{figure}[htbp]
  \centering
  \includegraphics{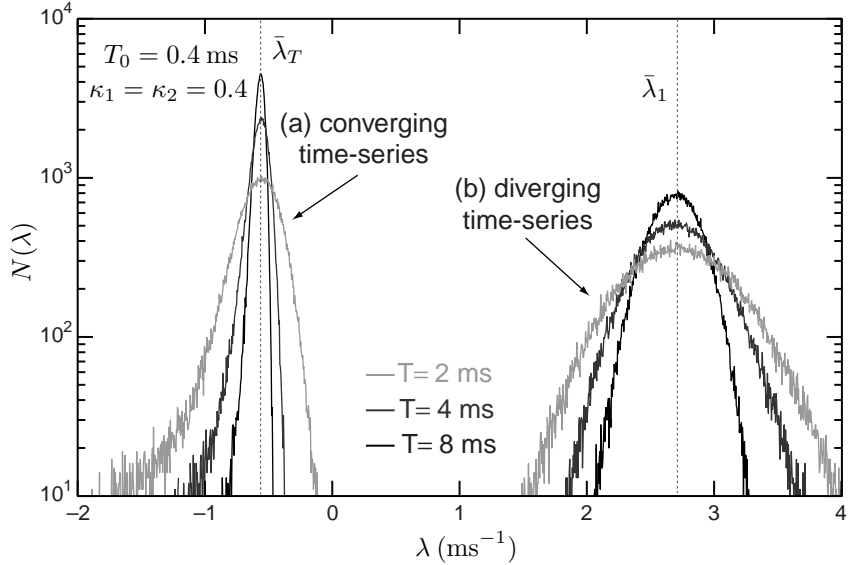}\\
  \caption{Histogram showing the distribution of finite-time transverse Lyapunov exponents, measured over time intervals of 2, 4, and 8 ms.  The transverse and maximal Lyapunov exponents were determined by numerically simulating the coupled system and fitting the convergence or diverence to an exponential relation, as depicted in figure~\ref{fig:convergence}(b) and (d).}\label{fig:fttle-histogram}
\end{figure}

In figure~\ref{fig:fttle-histogram}, we show distributions of local Lyapunov exponents for three choices of fitting time $T$ for $10^5$ simulated time-series.  The histograms labeled (a) show the distribution of transverse local Lyapunov exponents, obtained by simulating two initially independent systems that are suddenly coupled together with $\kappa_1 = \kappa_2 = 0.4$ at $t=0$.  The histograms labeled (b) histograms show the distribution of maximum Lyapunov exponents, obtained by simulating two initially synchronized systems that are suddenly released at $t=0$.  In all cases, the histogram is Gaussian near its peak and has non-Gaussian tails.  The mean of each distribution converges to the global or average (transverse) Lyapunov exponent $\bar\lambda$, and the standard deviation narrows in proportion to $T^{-1/2}$, as expected (Prasad \& Ramaswamy 1999, Ott 1993).

\begin{figure}[htbp]
  \centering
  \includegraphics{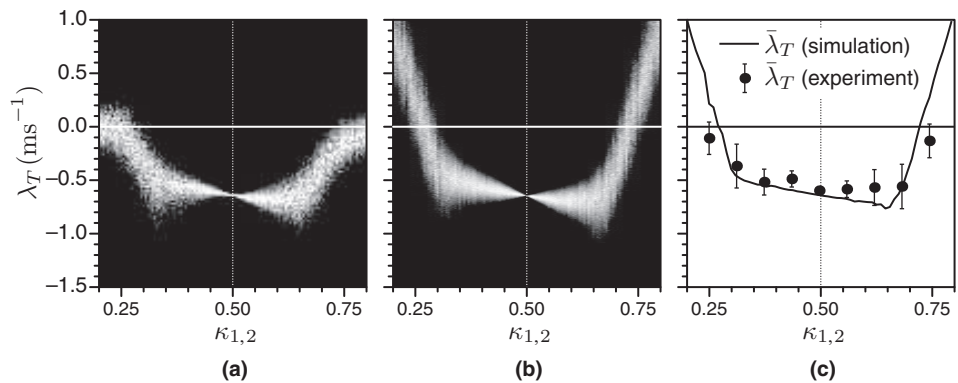}\\
  \caption{Distribution of transverse Lyapunov exponents as a function of the coupling strength $\kappa$, determined (a) by time-series analysis of converging transients and (b) by directly solving the linearized transverse equations of the coupled system.  (c) Comparison of calculated global transverse Lyapunov exponent and the measured finite-time transverse Lyapunov exponent, determined by measuring the transient convergence.  The data points and error bars indicate the average and standard deviation of the measured statistical distribution of $\lambda_T$, using a finite fitting time of 4 ms.}\label{fig:tle-vs-kappa}
\end{figure}

When a mathematical model of the system is available, the transverse Lyapunov exponents can also be calculated using the master stability function technique (Fujisaka \& Yamada 1983, Pecora \& Carroll 1998); i.e., by linearization about the synchronized chaotic solution.  Figure \ref{fig:tle-vs-kappa} compares the distribution of local transverse Lyapunov exponents obtained using both methods.  In figure~\ref{fig:tle-vs-kappa}(a) we plot (in grayscale) the distribution of local transverse Lyapunov exponents as a function of the coupling strength $\kappa$ $(=\:\kappa_1 = \kappa_2)$.  These histograms were obtained using time-series analysis to estimate the exponential convergence, as illustrated in figure \ref{fig:convergence}(b).  Because the time traces were initially uncorrelated, this method only applies when the TLE is negative, corresponding to convergent time series.  As anticipated from equation~(\ref{eq:s14}), the systems converge unconditionally when $\kappa_1 = \kappa_2 = 0.5$.  Figure \ref{fig:tle-vs-kappa}(b) plots the same distribution of TLEs, obtained by numerically solving the linearized system of equations.  Here, the linearized equations are sensitive to both positive and negative phase space growth, so the distributions can go above zero.  Apart from this expected difference, the correspondence between these two methods is remarkably good.

In figure~\ref{fig:tle-vs-kappa}(c), we plot the transverse Lyapunov exponents obtained from experimentally measured converging time-series of coupled systems.  At each value of $\kappa$, we measured the convergence rates $\lambda_T$ for 100 pairs of time-series.  The mean $\bar\lambda_T$ and standard deviation obtained by fitting the data to a Gaussian distribution are shown as the dots and bars on the figure respectively.  For comparison, the line indicates the `global' TLE computed using the linearized system of equations.  The experimental data agree well with the numerical simulations, which demonstrates that time-series analysis of converging experimental signals is a powerful technique for quantifying the transverse Lyapunov exponents of a system, even if a numerical model were unavailable.

\section{Adaptive synchronization}
\label{sec:adaptive}

As shown in \S\ref{sec:sync}, synchronization can depend on the coupling between oscillators.  In a practical network consisting of spatially separated chaotic oscillators, time-varying environmental conditions can cause the coupling to vary unpredictably.  Then, in order to maintain synchronization it is essential to dynamically compensate for these variations.  Recently, several algorithms have been developed to maintain or produce synchrony in a network of chaotic oscillators (Zhou \& Kurths 2006, Ito \& Kaneko 2001, De Lellis 2008, Feki 2003).  Sorrentino \& Ott (2008, 2009) proposed and simulated an adaptive algorithm to estimate and track a priori unknown coupling changes in a network of chaotic oscillators.  The estimate is then used to compensate for the environmental perturbations thereby ensuring synchrony.  In this section, we present an experimental demonstration of this scheme using a pair of nonlinear time-delayed optoelectronic feedback loops described in \S\ref{sec:oscillator}.  A DSP board, incorporated as part of the feedback loop (see figure~\ref{fig:dsp-setup}), enables us to perform real time computations allowing the implementation of the adaptive tracking algorithm.

\begin{figure}[htbp]
  \centering
  \includegraphics{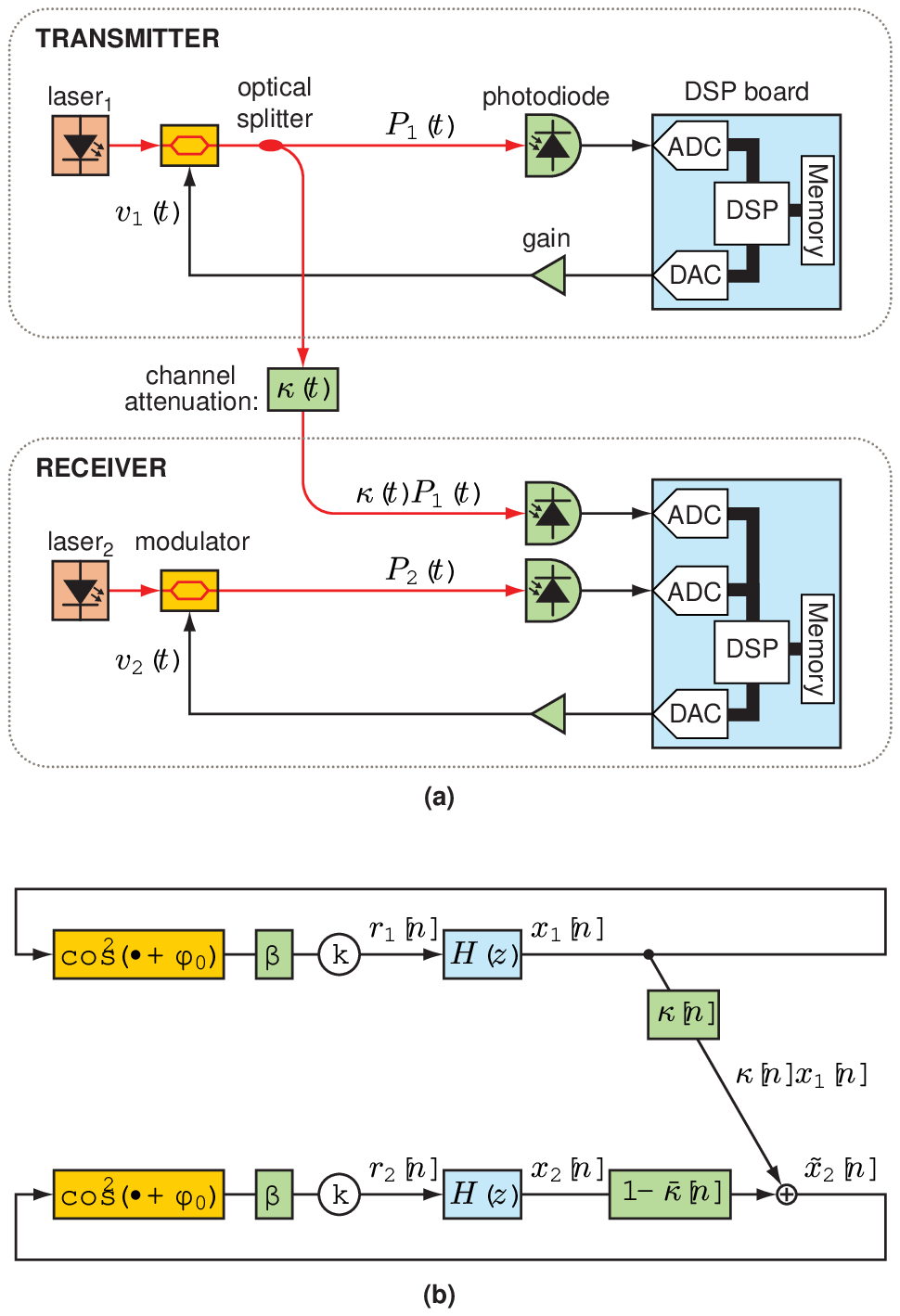}
  \caption{(a) Experimental setup of unidirectionally coupled feedback loops, in which the coupling factor $\kappa$ is allowed to vary slowly, to simulate the effect of an atmospheric perturbation or environmental disturbance.  (b) Equivalent discrete-time block diagram of two oscillators, unidirectionally coupled over a time-varying channel.  The receiver has no prior knowledge of $\kappa[n]$, and must therefore form an estimate, denoted $\bar\kappa[n]$ in order to keep the two systems sychronized.}\label{fig:unidirectional-coupling}
\end{figure}

In our experimental setup, shown in figure~\ref{fig:unidirectional-coupling}(a), we consider two optoelectronic feedback loops that are unidirectionally coupled through a time-varying communication channel, which is described by a coupling factor $\kappa(t)$.  An adaptive control scheme is implemented in the receiver in order to maintain synchrony between the two systems and, in the process, determine an estimate of the channel condition.

Figure \ref{fig:unidirectional-coupling}(b) shows an equivalent discrete-time mathematical block diagram of the two unidirectionally coupled systems with a time-varying channel.  Here we denote the channel coupling by $\kappa[n]$.  The receiving system has no a priori knowledge of $\kappa[n]$ and must therefore form an estimate, denoted $\bar\kappa[n]$, in order to maintain isochronal synchrony.  As before, the discrete-time bandpass filters $H(z)$ are governed by the state-space equations
\begin{align}
    \vv{u}_i[n+1] &= \vv{A}\vv{u}_i[n] + \vv{B}r_i[n] \\
    x_i[n] &= \vv{C}\vv{u}_i[n] + Dr_i[n] \\
    (i &= 1, 2) \nonumber\,,
\end{align}
where $r_i[n],\:(i=1,2)$ are the filter inputs, $x_i[n]$ are the corresponding filter outputs, and $\vv{A}$, $\vv{B}$, $\vv{C}$ and $D$ describe the bandpass filter.  The filter outputs are fed back to the inputs through a nonlinearity and time delay according to
\begin{align}
    r_1[n] &= \beta\cos^2(x_1[n-k] + \phi_0) \\
    r_2[n] &= \beta\cos^2\bigl[(1-\bar\kappa[n-k])x_2[n-k] + \kappa[n-k]x_1[n-k] + \phi_0\bigr]\,,
\end{align}
where $\bar\kappa[n]$ is the local estimate of the channel coupling.

One can clearly see that these equations admit a synchronous solution in the case that $\bar\kappa[n] = \kappa[n]$, i.e., provided the receiver tracks the coupling strength $\kappa[n]$.  The analysis presented in \S\ref{sec:coupling} showed that for static coupling, i.e., when $\kappa[n] = \bar\kappa[n] \equiv \kappa$, the synchronous solution is stable over a continuous range of values of $\kappa$.  This result suggests that if $\kappa[n]$ varies slowly, while remaining within the bounds required for synchrony stability, the systems could stay synchronized as long as the receiver is able to track the variation with sufficient accuracy.  We emphasize that the receiver does not have direct knowledge of $\kappa[n]$ but only receives the product $\kappa[n]x_1[n]$, as shown in figure \ref{fig:unidirectional-coupling}(b).

Sorrentino \& Ott (2008, 2009) prescribed a strategy in which the local factor $\bar\kappa[n]$ is adjusted in a way that minimizes the average synchronization error.  This yields the following estimate $\bar\kappa[n]$,
\begin{equation}\label{eq:as5}
    \bar\kappa[n] = \frac{\left<\kappa x_1x_2\right>_{\rm LPF}}{\left<x_2^2\right>_{\rm LPF}} \equiv \frac{N[n]}{D[n]}\,,
\end{equation}
where $\left<\bullet\right>_{\rm LPF}$ denotes an exponentially-weighted moving average, which is equivalent to a discrete-time low-pass filter.  This averaging process can be implemented with the following discrete-time iterative equations:
\begin{align}
    N[n] &= z_0 N[n-1] + (1-z_0)\kappa[n] x_1[n]x_2[n]\\
    D[n] &= z_0 D[n-1] + (1-z_0)x^2_2[n]\,,
\end{align}
where the forgetting factor $z_0$ is the pole of the discrete-time low-pass filter.  The time-window over which the averaging is performed is approximately $T_s(1-z_0)^{-1}$, where $T_s$ is the sampling period.  We note that, as required, the adaptive scheme described by equation (\ref{eq:as5}) relies only on the product $\kappa[n]x_1[n]$ and $x_2[n]$ to form the estimate $\bar\kappa[n]$.  In a high-speed application, the lowpass filter could easily be implemented using an electrical mixer in place of discrete-time averaging filter.

We experimentally demonstrated the adaptive synchronization scheme using a pair of coupled nonlinear optoelectronic oscillators, as shown in figure \ref{fig:unidirectional-coupling}(a).  For these experiments, the bandpass filters were adjusted to have a passband of 100 Hz -- 2.5 kHz, the DSP sampling frequency was reduced to 24 kS/s, and the time delay was measured to be $k = 36$ timesteps, or 1.5 ms.  We chose a feedback strength of $\beta = 3.58$, which, under these conditions, was found to yield robust chaotic behaviour.  The lowpass filter used in the adaptive synchronization rule was implemented with a forgetting factor of $z_0 = 0.95$, which corresponds to a filter response time of 208 $\mu$s.

\begin{figure}[htbp]
  \centering
  \includegraphics{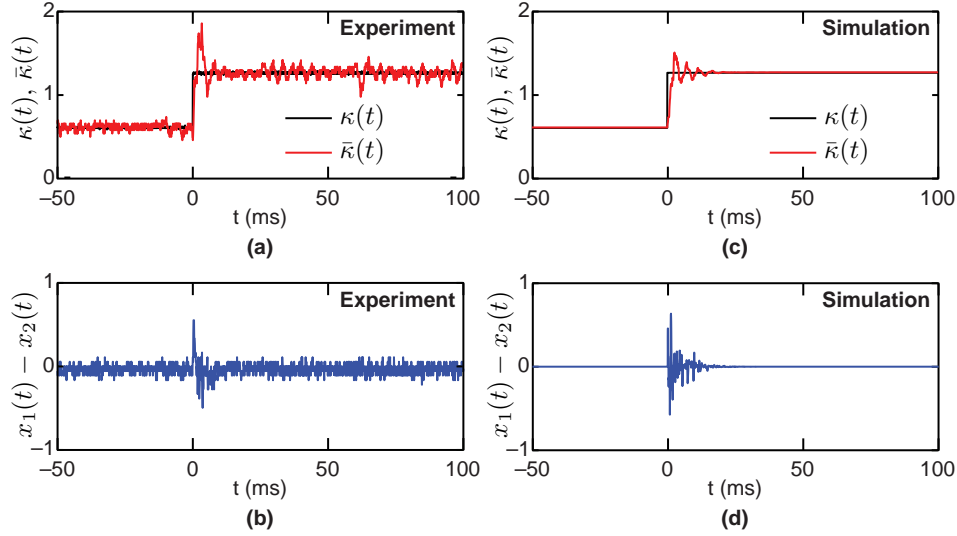}
  \caption{(a) Measured and (b) simulated response of adaptive coupling system to a sudden change in $\kappa$.  In these experiments, the coupling strength $\kappa$ was changed abruptly from 0.80 to 1.13 at $t=0$.  The adaptive synchronization scheme automatically adjusts $\bar\kappa$ in response to this variation.  Here we plot both the tracking signal $\bar\kappa(t)$ and the difference $x_1(t) - x_2(t)$, showing the initial loss of synchrony followed by recovery.}\label{fig:adaptive-sync}
\end{figure}

Figure~\ref{fig:adaptive-sync} presents experimental measurements and numerical simulations showing how both the synchronization error and tracking signal $\bar\kappa[n]$ respond to an abrupt change in the coupling from $\kappa = 0.8$ to $\kappa = 1.13$.  For $t<0$ the coupling strength $\kappa$ was held constant at $\kappa = 0.80$.  Under these conditions the receiver forms the correct estimate $\bar\kappa = 0.8$, which gives a small synchronization error.  At $t=0$ the coupling strength was switched abruptly to  $\kappa = 1.13$, which causes the two loops to briefly lose synchrony.  However, the receiver adaptively readjusts the parameter $\bar\kappa$  to track $\kappa[n]$ and the synchrony is regained.  The numerical simulations shown in figure \ref{fig:adaptive-sync}(c) and (d) exhibit similar behaviour.  The response time of the adaptive synchronization method was found to be limited primarily by the exponentially-weighted moving average filter.  In a separate work we studied the ability of this adaptive scheme to track sinusoidal variations in coupling, and we quantified the limitations on the magnitude and frequency of the perturbation that can be tracked (Ravoori \etal 2009).

\section{Conclusion}
\label{sec:conclusion}

Many potential applications such as secure communication, sensor networks, spread-spectrum communication, chaotic radars and random number generators could benefit from a nonlinear dynamical system that is simple to model, easy to implement, and capable of generating robust, high-dimensional, chaotic waveforms.  This paper presents a comprehensive analysis and characterization of a nonlinear optoelectronic feedback system that meets these criteria.  The system uses electrooptic modulation and optical transmission, and it can therefore take advantage of the vast array of low-cost, high-speed, widely available components originally developed for fibre-optic communication networks.  We describe a new approach in which the delayed electrical feedback and filtering is implemented using real-time digital signal processing.  This greatly facilitates matching of filter characteristics between systems, and also allows for real-time control and adjustment of the feedback parameters -- something that could not be easily accomplished with traditional analogue signal processing.

Because most of the aforementioned applications of chaotic signals require synchronization between two or more systems, we explore the conditions under which coupled system will synchronize.  We present a new technique to experimentally quantify the rate of convergence when two systems are coupled and the rate of divergence when they are released.  Finally, we demonstrate an adaptive technique that automatically maintains synchronization between coupled systems, in the presence of an unknown and time-varying coupling between the two.

\begin{acknowledgements}
This work was supported by DOD MURI grant (ONR N000140710734) and the US-Israel Binational Science Foundation.
\end{acknowledgements}

\label{lastpage}
\end{document}